\documentclass{article}

\usepackage{amsmath}

\usepackage{amssymb}
\usepackage{amsfonts}
\usepackage{graphicx}
\usepackage{doi}
\usepackage{cite}
\newcommand\eps{\varepsilon}

\begin{document}

\title{Intransitiveness in the Penney Game and in Random Walks on rings, networks, communities and cities}
\author{Alberto Baldi$^{1}$, Franco Bagnoli$^{1,2}$\\
$^{1}$ \quad Department of Physics and Astronomy and CSDC, \\
University of Florence, \\
via G. Sansone 1, 50019 Sesto Fiorentino, Italy.;\\ alberto.baldi1@stud.unifi.it, franco.bagnoli@unifi.it\\
$^{2}$ \quad Also INFN, sez. Firenze. }

\maketitle

\abstract{The concept of intransitiveness for games, which is the condition for which there is no first-player winning strategy can arise surprisingly, as happens in the Penney game, an extension of the heads or tails.  Since a game can be converted into a random walk on a graph, i.e., a Markov process, we  extend the intransitiveness concept to such systems. The end of the game generally consists in the appearance of a pre-defined pattern. In the language of random walk this corresponds to an absorbing trap, since once that the game has reached this condition the game  comes to an end. Therefore, the intransitiveness of the game can be mapped into a problem of competition among traps. We analyse in details random walkers on several kind of  networks (rings, scale-free, hierarchical and city-inspired) with several variations: traps can be partially absorbing, the walker can be biased and the initial distribution can be arbitrary. 
We found that the transitivity concept can be quite useful for characterizing the combined properties of a graph and that of the walkers.}


\catcode`\@=\active
\def@#1{\boldsymbol{#1}}

\section{Introduction}
  
Games are an integral part of all cultures and are one of the oldest forms of social interaction. Many games use stochastic elements (dice, coins, wheels) and most of them are based on betting for the occurrence of a given pattern, which can be just a single number (like in Roulette, Head or Tail) or more complex configurations like in card games. 

It is possible to formalize some of the simplest games as a random walk on the graph of all possible patterns, so that the victory corresponds to reaching a given node of the graph. 

Although some games are completely determined by the sequence of aleatory events, like in the Game of the Goose, in others the player may employ a strategy (or choose the target) so to increase his/her chance of success. For instance, in the game of betting on the sum of two dice (dice roulette), the strategy of betting on the seven is surely advantageous, in statistical terms.

In many such games there exists an optimal strategy, generally for the first player. In the dice roulette, if two players are not allowed to bet on the same exit number, the first player is surely favoured. 

A game with hierarchy of strategies (which can be just a move) is called transitive: if strategy A beats strategy B and B beats C, then A beats C. 
There are also non-transitive game. Typical example is the game rock-scissors-paper: rock beats scissors, scissors beats paper, paper beats rock.
  
In this article we shall start exploring an intransitive game, called the Penney game~\cite{Penney}, which  provides a general and simple definition of "globally intransitive" systems. The Penney game can be reformulated as a random walk  on a specific network, where the nodes represent the possible choices which the players can bet on. When the betting choices have been made, the selected nodes become targets and can be represented by absorbing traps for the random walker.
  
In our analysis we shall observe that the intransitiveness property is strictly related to the directionality of the graph of the random walk. 
This will bring us to explore the simplest example of directed network: a one-dimensional lattice with periodic boundary conditions.
  
The network topology in general plays a key role in defining  the strategic positioning of the trap nodes in order to capture the most of random walkers.
Moreover, the intransitiveness of a system is related not only to the topology of the network that we use to schematize it, but also to absorbency of traps, i.e., the ability of players to recognize their victory. For this reason we shall also explore the effect of partially absorbing traps. 

These first two models are at the same time simple and representative of the main characteristics of intransitive games. However, the devised formalism is general and we only need the adjacency matrix of the network to fully describe the competitive behaviour. Consequently, it can be applied to many different cases, from competition for the best positioning of a shop in a urban network, to the visibility of sites in the Internet. We explored this last case in the last section, where we analyse scale-free, hierarchic and city-like networks.

\section{Intransitiveness}
\label{global_intransitiveness}
Binary relations are often characterized by transitivity, a property of a relation $\sim$ defined on a set $\mathcal{I}$, such that 
\[
A\sim B, B \sim C \Rightarrow A\sim C \hspace{.5 cm} \forall A,B,C \in \mathcal{I},
\] 
and a relation is not transitive if this condition is not valid for at least a triple. Instead, a relation is intransitive when
\[
A \sim B, B \sim C \Rightarrow \neg (A\sim C) \hspace{.5 cm} \forall A,B,C \in \mathcal{I},
\]
where "$\neg$" denotes the negation (NOT).

Many games are based on transitive rules, but there are also games with intransitive rules. This is the case of the rock-paper-scissors game: rock wins against scissors, which in turn wins against paper, which in turn wins again against rock. Games like this one are defined "globally intransitive" and are characterized by the fact that, if a player knows the strategy of his/her opponent, he/she has a higher probability of winning by choosing the optimal strategy.

In order to quantify the intransitiveness degree of a system, we introduce the victory matrix $V$ and the index of global intransitiveness $\sigma$. The first one is defined as a square matrix of all possible choices of  players (three for the rock-scissors-paper game) and the generic entry $V_{ij}$ is the probability that choice $i$ will win against choice $j$. Notice that this matrix is for the second player.

The index $\sigma$ is defined as:
\begin{equation}
  \label{def sigma}
  \sigma = \min_{i = 1 \, ... \, N} \left \{\max_{j= 1 \, ... \, N} V_{ij} \right \} -\frac{1}{2},
\end{equation}

The quantity  $\max_{j= 1 \, ... \, N} V_{ij}$ for a specific $i$ corresponds to the maximum winning probability given the $i$ move. Then, taking the minimum over index $i$ we find the worst case, i.e. the least winning probability for the player. In order to have global intransitiveness this quantity needs to be larger than one half, so that, even in the worst case, the player is favoured.



Summarizing:
    if $\sigma > 0$, the system is globally intransitive and the choice of the first agent can be countered by an optimal choice of the second, who statistically wins; if $\sigma = 0$, the two players can "tie", because the system permits at least two equivalent strategies;
    if $\sigma < 0$, the system is transitive and so, if the first agent makes the optimal choice, he/she statistically wins against the second. 

It is possible to reformulate a game as random walk on a directed network where the nodes represent the possible choices of the players and the links describe the competitive relations. Let us illustrate it for the Penney game.

\section{The Penney game}
This game is an extension of coin toss game, in which two players are asked to bet not on a single coin flip, but on a contiguous succession of heads and tails, whose length $l$ is fixed. This gives to each player a choice among $2^l$ different sequences of heads and tails. The game is won by the player whose sequence will be extracted first.

\begin{figure}[t]
  \begin{center}
    \begin{tabular}{c c}
      \includegraphics[scale=0.7]{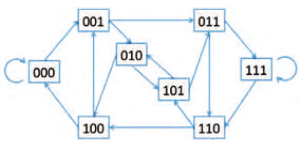}  &
      \includegraphics[scale=0.4]{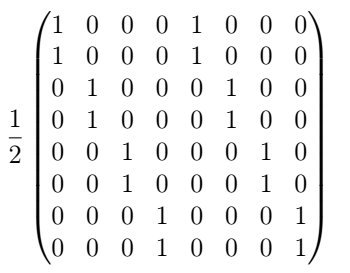} \\
      (a)  & (b) \\
      \begin{minipage}[b]{0.45\columnwidth}\small
        \[
        \left(\begin{matrix}
        0 & \frac{1}{2} & \frac{2}{5} & \frac{2}{5} & \frac{1}{8} & \frac{5}{12} & \frac{3}{10} & \frac{1}{2} \\[.2cm]
        \frac{1}{2} & 0 & \frac{2}{3} & \frac{2}{3} & \frac{1}{4} & \frac{5}{6} & \frac{1}{2} & \frac{7}{10} \\[.2cm]
        \frac{3}{5} & \frac{1}{3} & 0 & \frac{1}{2} & \frac{1}{2} & \frac{1}{2} & \frac{3}{8} & \frac{7}{12} \\[.2cm]
        \frac{3}{5} & \frac{1}{3} & \frac{1}{2} & 0 & \frac{1}{2} & \frac{1}{2} & \frac{3}{4} & \frac{7}{8} \\[.2cm]
        \frac{7}{8} & \frac{3}{4} & \frac{1}{2} & \frac{1}{2} & 0 & \frac{1}{2} & \frac{1}{3} & \frac{3}{5} \\[.2cm]
        \frac{7}{12} & \frac{3}{8} & \frac{1}{2} & \frac{1}{2} & \frac{1}{2} & 0 & \frac{1}{3} & \frac{3}{5} \\[.2cm]
        \frac{7}{10} & \frac{1}{2} & \frac{5}{8} & \frac{1}{4} & \frac{2}{3} & \frac{2}{3} & 0 & \frac{1}{2} \\[.2cm]
        \frac{1}{2} & \frac{3}{10} & \frac{5}{12} & \frac{1}{8} & \frac{2}{5} & \frac{2}{5} & \frac{1}{2} & 0 \\
        \end{matrix}\right) 
        \]\\[-.4cm]
        
      \end{minipage} 
      &
      \begin{minipage}[t]{0.45\columnwidth}
        \includegraphics[width=\columnwidth]{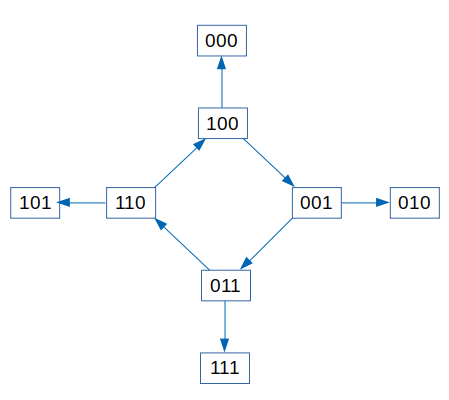}  
      \end{minipage}\\
      (c) & (d) \\
    \end{tabular}
  \end{center}
  \caption{Quantities for the Penney game with $l=3$. (a) Transition graph, every node is a possibly winning sequence, the links go from one sequence to the sequences which can be obtained with a coin toss. Each link weights 1/2.
    (b) Weighted adjacency matrix (Markov matrix) $@M$ of the system, where the  sequences (indexes) are read as base-two numbers.
    (c) Victory matrix and  (d) victory graph.}
  \label{fig:grafopenneyl3}
\end{figure}

Every possible sequence has the same probability to appear in the first $l$ coin tosses, and considering infinite tosses every succession appears the same number of times. These observations might suggest that betting on a sequence rather than on another is completely equivalent but this is not true: once the first sequence of $l$ coin tosses has been extracted, the second one has the first $l-1$ symbols in common with the first one, and so on for the following sequences. For any extraction there exist only two possibilities for the following sequence. This ultimately implies that there exist specific paths in the space of the $2^l$ sequences that one is forced to follow and the starting sequence (which is actually random) represents the initial condition which affects all the following paths. The Penney game can be therefore represented as a random walk process on a directed network with $2^l$ nodes, where the directed links connect sequences which can appear one after the other. The coin extraction process is therefore figuratively represented by the moves of a random walker which travels the network by following the directed links.
In Fig.~\ref{fig:grafopenneyl3} the network and the relative adjacency matrix for the case $l=3$ are shown.


\section{Partially absorbing traps and biased walks}
This change of perspective allows us to extend the following analysis also to  other competitions situations, like shops competing for customers. In this case we can think of the network given by streets, approximating customers as random walkers looking for a certain product. If they enter the first shop selling that product. shops act as absorbing traps, and an interesting question is whether, given a city map, there exists locations that are robust with respect to competing shops ($\sigma <0$) or not ($\sigma >0$)~\cite{fanelli}. 

Another example is that of the searches by web crawlers. A robot is downloading web pages searching for some piece of information, by navigating the links (and eventually jumping to other pages, as in the Google algorithm~\cite{brin,langville}). This again is an example of a random walk on a graphs (the web network) with absorbing traps (the information) and one could be interested in where to place it in order to maximise the probability of being found, while minimizing that of being obscured by another page. 

It is also interesting to study the effect of partially absorbing traps, which corresponds to distracted players, who do not realize that the winning combination has appeared, or, more realistically, to customers which are not fully satisfied by a given product, which however can be taken into consideration. We shall indicate with $\eps$ the absorbency of traps ($\eps =1$ for fully absorbing traps). 

Finally, one can be interested in studying the case of biased walks, which correspond to biased coins in the Penney game, or to biased walks in towns, and weighted jumps for web crawlers. The biasing parameter will be indicated by $p$ ($p=1/2$ for unbiased paths). 

Therefore, studying $\sigma$, we can characterize the competition in systems describable as random walk on network.

\begin{figure}[t]
  \begin{center}
    \begin{tabular}{cc}
      \includegraphics[width=0.45\columnwidth]{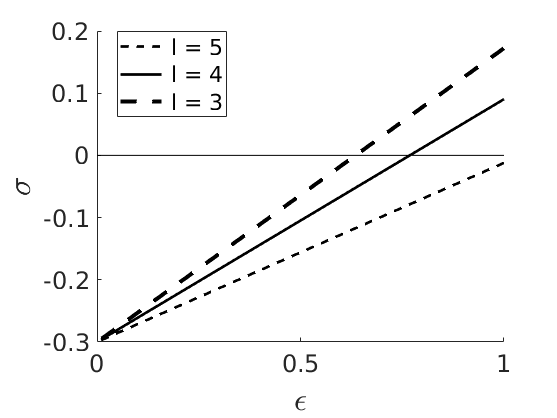}& 
      \includegraphics[width=0.45\columnwidth]{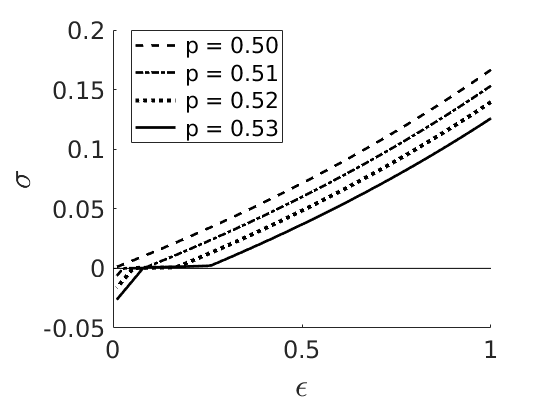}\\
      (a) & (b) \\
    \end{tabular}
    \caption{(a) Index of global intransitiveness $\sigma$ as function of the absorbency $\eps$, for fixed bias $p = \frac{1}{5}$, for sequences of different length $l$.       (b) Index of global intransitiveness $\sigma$ as function of $\eps$, fixed $l=3$ for several values of $p$.}
    \label{penneypfixedl3}
  \end{center}
\end{figure}

\section{Random walks, traps and the victory matrix}

Let us consider a random walk process on a  network, which can be described as 
\begin{equation}\label{M}
\omega_i(t+1)=\sum_j\frac{A_{ij}}{k_j}\omega_j\equiv\sum_jM_{ij}\omega_j,
\end{equation}
where $\omega_i(t)$ is the probability of finding a walker on node $i$ at time $t$, $A$ is the adjacency matrix, $k_j=\sum_i A_{ij}$ is the out-degree of a generic node $j$ and $M$ is resulting  stochastic matrix of our Markovian process.

The victory matrix element $V_{ij}$ is given by the number of walks arriving at node $i$ without passing through node $j$. The simplest way to compute it is to consider nodes $i$ and $j$ as traps, so that we are sure that, after a fairly large number of temporal steps, every possible path of the casual agent will end in one of the two nodes. 

We have to keep on mind that the victory matrix also depends on the  initial distribution $@\omega(0)$. While for games like the Penney one it is natural to start from a uniform distribution (at beginning no coin is shown), there are variations (like the Texas Hold'em poker) in which one has a previous information. Also for walkers in city and web crawlers, it may happen that the starting point is predefined. 

\subsection{Victory matrix  for fully absorbing traps}

In order to compute $V_{ij}$ for fully absorbing traps, it is  necessary to count the number of paths ending at $i$ and avoiding $J$ . 
This distribution of path ending a given site, $@\omega(t)$, evolves over time by the iteration of a modified stochastic matrix, which takes into account the presence of the absorbing nodes $i$ and $j$,
\begin{equation}
\label{evol_trap}
@\omega(t+1) = @M^{[i,j]} @\omega(t),
\end{equation}
where  $@M^{[i,j]})$ is obtained modifying the stochastic matrix $@M$ of Eq.\eqref{M}, by setting the elements of columns $i$ and $j$ equal to zero, except the diagonal elements $(M^{[i,j]})_{ii}$ and $(M^{[i,j]})_{jj}$, which are set equal to one so that, when the casual walker arrives on nodes $i$ and $j$, he/she stays there forever.

We can introduce the trap matrix $T^{[ij]}$ with is zero except for $(T^{[i,j]})_{ii}=(T^{[i,j]})_{jj}=1$, so that
\[
@M ^{[i,j]} = M (\mathbb{I}-T^{[i,j]}) + T^{[i,j]}.
\]

let us denote by  $\tilde{@\omega}=\lim_{t\rightarrow\infty}@\omega(t)$.If the matrix $@M$ is not singular (all nodes are connected),  in the long-time limit only $\tilde{@\omega}_i$ and $\tilde{@\omega}_j$ are non-zero. 
Therefore, we can define the victory matrix $V_{ij}$ as
\[
V_{ij} = \frac{\tilde{\omega}_i}{\tilde{\omega}_i+\tilde{\omega}_j}.
\]

%

\subsection{Victory matrix  for partially absorbing traps}
In the case of partially absorbing traps, we need to modify the algorithm, using two distributions, $@{\omega}$ as above for the walkers that stay on the lattice (not fallen in the traps) and $@\tau$ for the traps. 
Their evolution is given by 
\[
\begin{split}
@\omega(t+1) =&@M(\mathbb{I}-\eps @T^{[i,j]})@\omega(t),\\
@\tau(t+1) =&@\tau(t) + \eps @T^{[i,j]}@\omega(t),
\end{split}
\]
where $\eps$ is the absorptivity of traps.

\section{Numerical results}
\label{Results}

\subsection{Intransitiveness in the Penney Game}
Let us consider the Penney Game for sequences of length $l \ge 3$ (Fig.~\ref{fig:grafopenneyl3}).
Every column of the Vicory matrix has an element bigger than $1/2$, and this means that every sequence is statistically beaten by at least another one. This indicates that the Penney game for $l=3$ is  a globally intransitive system, and this is true also for larger sequence lengths. The same holds for any $l>2$, the only exception is the case $l=2$, where $\sigma = 0$, because the sequences 10 and 01 have the same winning probability against each other.


When a system is globally intransitive, we can illustrate this property using a ``victory graph'', considering the maximum value that appears in every column of the victory matrix (therefore, for every sequence we consider the one that has the greatest chance of winning against it), Fig.~\ref{fig:grafopenneyl3}-d. In this case, arrows have a precise meaning: sequences from which the arrows start are those that most likely to beat the arrival sequences.  In globally intransitive systems the victory graph is always connected and it is characterized by a loop of nodes at centre.

Let us now investigate how this situation modify if we consider less absorbing traps ($\eps<1$) or a biased coin $p\neq 1/2$).  Assuming the same level of distraction for the two players (the two absorbing traps have same degree of absorbency), we note how, studying the curve of $\sigma$ when varying $\eps$, we can identify regions where $\sigma <0$ (transitivity), $\sigma = 0$ (tie) and other where $\sigma > 0$ (intransitiveness).

As shown in Fig.~\ref{penneypfixedl3}-a , we note that, generally, the index of global intransitiveness  $\sigma$ is an increasing function of the absorbency parameter $\eps$. When the system is not globally intransitive ($\sigma < 0$), as $\eps$ increases, the transitivity degree of the system grows. The length size  $l$ also plays a key role for the intransitiveness:  completely transitive systems, as the Penney game with $p = \frac{1}{5}$, can show a global intransitiveness zone for high values of $\eps$ by increasing the length $l$ of the sequences.

Fixing instead the size of the system (Fig.~\ref{penneypfixedl3}-b), studying the behaviour of $\sigma$ in the neighbourhood of $\frac{1}{2}$, we observe how the system can be globally intransitive $\forall \, \eps$ only when we consider the fair Penney Game, while for biased coin, departing from $p = \frac{1}{2}$, the transitivity zones are always present for small values of the traps absorbency. Clearly, when only heads or tails can come out as results of the coin tosses, the system is fully transitive and $\sigma = -\frac{1}{2}\, \forall \, \eps$.

\begin{figure}[t]
\centering
        \begin{tabular}{ccc}
\includegraphics[width=0.32\linewidth]{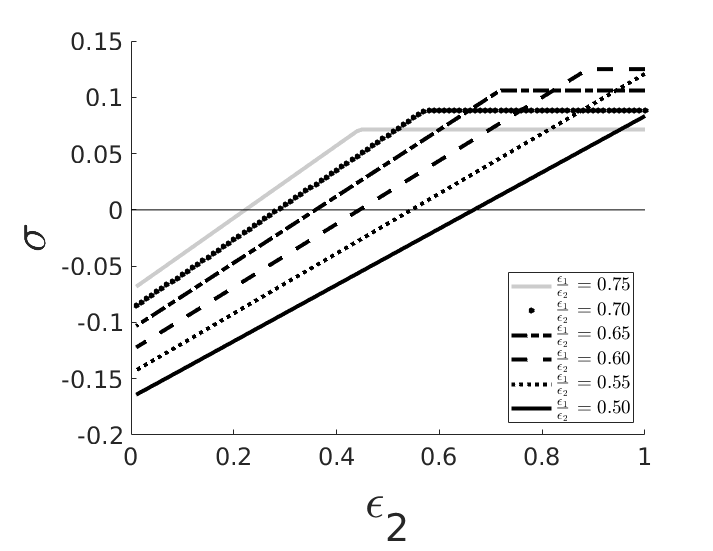} &
\includegraphics[width=0.32\linewidth]{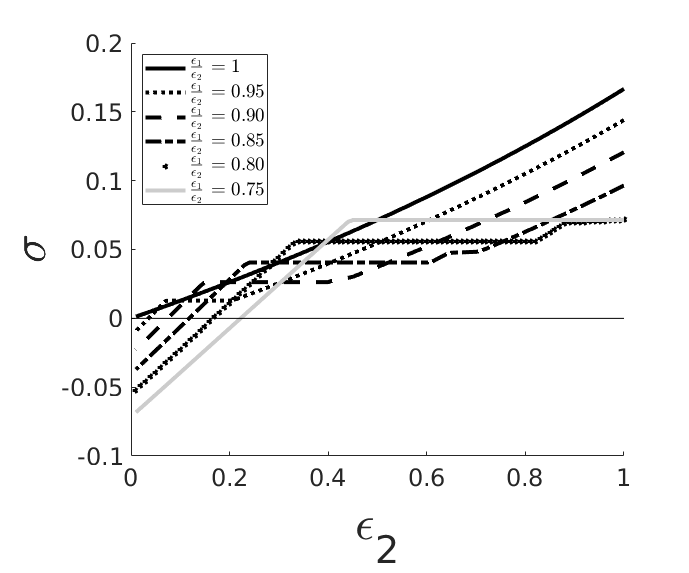} & \includegraphics[width=0.32\linewidth]{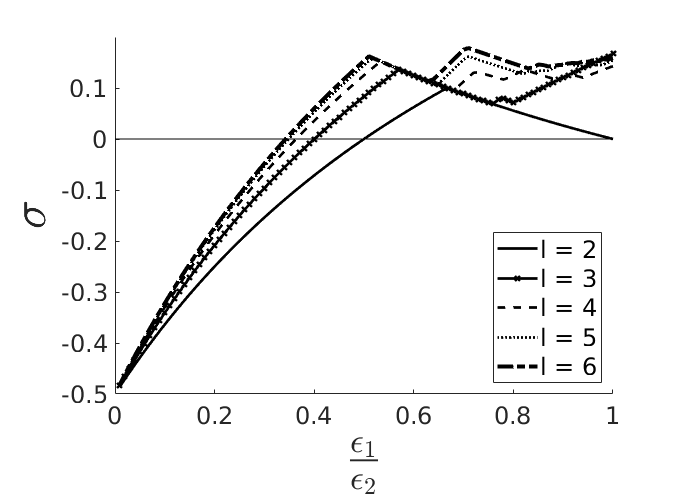}\\
          (a) & (b) &(c)\\
\end{tabular}
\caption{The intransitiveness index $\sigma$ for (a) various values of the absorbency of the two traps. One can notice that the intransitiveness ($\sigma>0$) arises only for absorbency of traps sufficiently high. (c) Similar scenarios show for $l>3$, while one can notice that for $l=2$ we have intransitiveness for non-symmetric traps ($\epsilon_1/\epsilon_2<1$). A similar effects happens for $l=2$ and biased walks $p\neq 1/2$ (not reported here).}
\label{fig:dfferenttraps}
\end{figure}

It is also interesting to investigate the case in which the absorbency of traps is not equal, or, in other words, one of the players is more attentive than the other, results are reported in Fig.~\ref{fig:dfferenttraps}.

\subsection{Intransitiveness and competition on cycles}
The Penney Game victory graph of Fig.~\ref{fig:grafopenneyl3} shows a cycle inside its structure. It is therefore interesting to investigate the "core" of that graph, the cycle itself. 

Let us investigate the competition of traps located on a cyclic graph, with walkers that move in a given direction with probability $p$. As in the Penney Game case, we consider the same level of absorbency for the traps. Generally, we expect  competition on a directed cycle with $p=0$ or $p=1$ to be the globally intransitive system par excellence: for each node of the network where a trap can be located, there are other nodes (the immediately preceding neighbours, considering the direction in which the agent moves), that can "obscure" it. As an applied example, consider a rotatory square (with periodic boundary conditions) where two or more shops selling the same product compete for customers.
The higher the quality and the price of goods in one shop, the higher its appeal, and consequently its absorbency $\epsilon$.

\begin{figure}[t]
    \begin{center}
        \begin{tabular}{cc}
            \includegraphics[width=0.45\columnwidth]{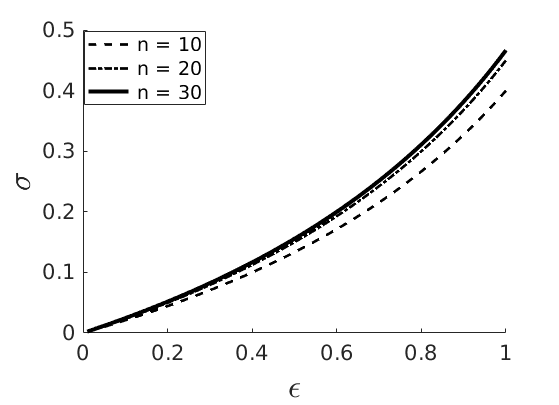} & 
            \includegraphics[width=0.45\columnwidth]{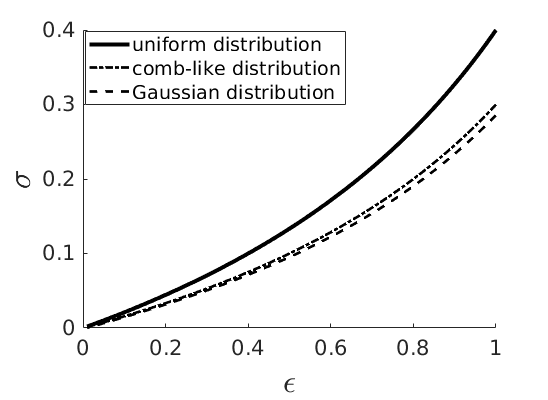} \\
            (a) & (b)\\
             \includegraphics[width=0.45\columnwidth]{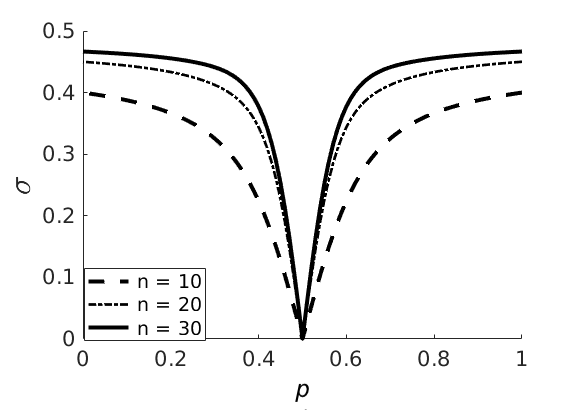}  & 
            \includegraphics[width=0.45\columnwidth]{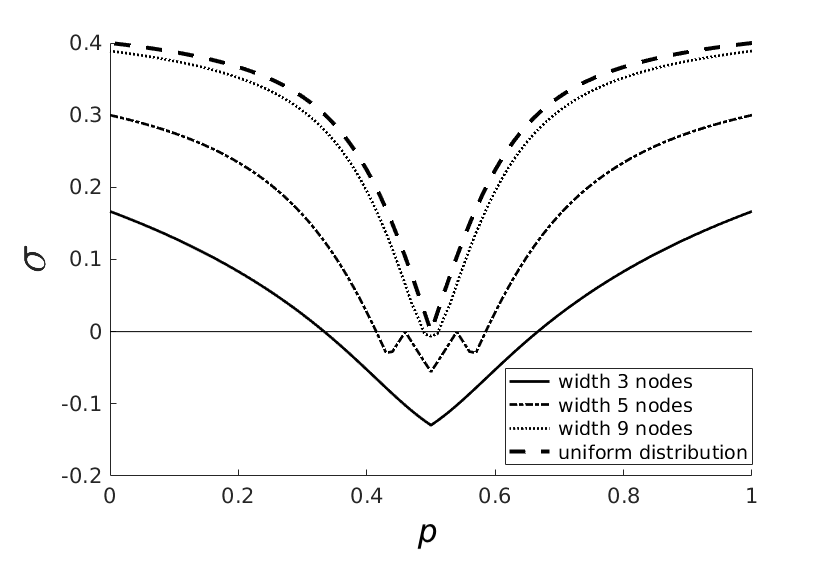} \\
            (c) & (d) \\
        \end{tabular}
        \caption{(a) Intransitiveness index $\sigma$ as function of $\epsilon$ for a cycle, for different sizes of the system and assuming an initial uniform probability distribution. (b) Intransitiveness index $\sigma$ as function of $\epsilon$ for a cycle with 10 nodes. A different initial distribution of probability is associated to every type of line. Gaussian distribution has a standard deviation equal to 0.15.
          (c) Intransitiveness index $\sigma$ as function of $p$, fixed the size of the system and assigned an initial uniform distribution of probability. (d) Intransitiveness index $\sigma$ as function of $p$, for a cycle of 10 nodes, assigned an initial step-like distribution of probability. A different width of the step is associated to every type of line.
            \label{fig:distr}}
    \end{center}
\end{figure}

Let us analyse the competition between two equivalent (same value of $\eps$) traps on the cycle, initially assuming a uniform probability distribution for the casual agents on the network.
Similarly to the Penney Game, we notice that also the one-dimensional lattice is globally intransitive $\forall \, \eps$, and that increasing the size of the network increases the intransitiveness degree, see Fig.~\ref{fig:distr}-a.

Moreover, if we consider fully absorbing traps ($\eps=1$), the intransitiveness $\sigma$ tends to $\frac{1}{2}$  when increasing the size of the system, which corresponds to the case where for each position of a trap on the network, there is at least another position that is able to statistically obscure it.

A similar behaviour of $\sigma$ is  also observed considering different initial probability distributions, assuming that a zone of the cycle is more populated in a neighbourhood of a specific node by the walkers. This can be represented by a Gaussian distribution or by a comb-like distribution, in which nodes with a constant initial density of walkers are alternated with nodes with no walkers. 

For high values of $\eps$, considering these initial distribution, we still observe positive values for $\sigma$, but smaller in amplitude (Fig.~\ref{fig:distr}-b). The values taken by the index of global intransitiveness decrease because the traps can be located in nodes where the initial density of walkers is higher. For instance, let us compare the comb-like distribution with the uniform one in the case of fully absorbing traps located alternated one by one on the network.

\begin{figure}[t]
    \begin{center}
        \begin{tabular}{ccc}
            \includegraphics[width=0.32\columnwidth]{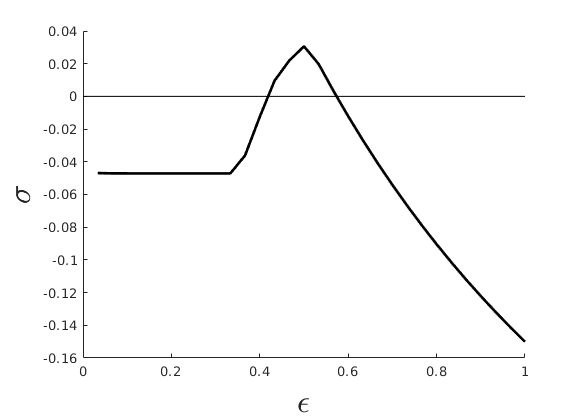} &
            \includegraphics[width=0.32\columnwidth]{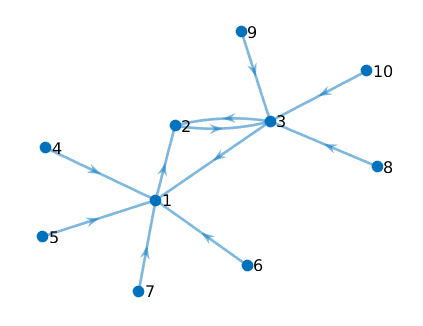} &
            \includegraphics[width=0.32\columnwidth]{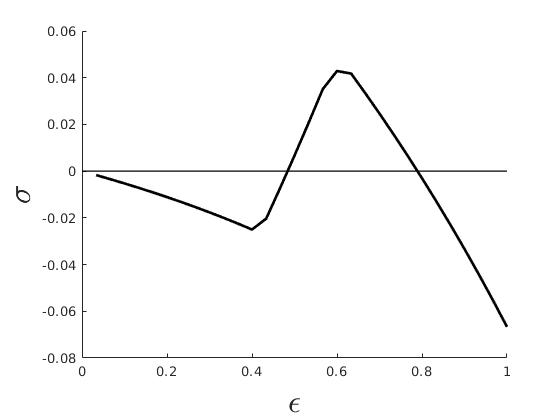} 
            \\
            (a) & (b) & (c) \\
        \end{tabular}
        \caption{(a) Intransitiveness index $\sigma$ vs absorbency $\epsilon$ for a scale-free network with 100 nodes as a function of the absorbency parameter $\eps$. There is an intransitive behaviour for intermediate values of this parameter. (b) 
          The small subnet presumably responsible for the  the intransitive region  of $\sigma$ vs $\eps$. (c) Intransitiveness index $\sigma$ vs absorbency $\epsilon$ for this  subnet.
          \label{fig:net}
        }	
    \end{center}
\end{figure}

Since $\eps = 1$, considering two traps located at distance two, in two nodes where the density of walkers is identical and non-zero, the first one (in the sense of the cycle) obscures the second one almost completely, catching also all the other incoming walkers. This corresponds to the disposition of traps that allows to one trap the maximum winning probability against any other trap. The second trap in this case only catches the walkers which are already present here, who are more numerous in the case we consider a comb-like distribution than an uniform. This  causes the diminution of the amplitude of $\sigma$ for the comb-like distribution, with this effect that decreases for $\eps \rightarrow 0$.

The curve of $\sigma$ considerably changes if all the walkers are initially placed in a single node. In this case the system is always transitive, regardless of the absorbency of the traps: if one of the latter is located in the node initially characterized by the presence of all the walkers, even if weak, it always exerts a greater attraction for them,  who are less likely to be attracted by the other nodes of the network, in which the opposing trap can be located.
Clearly, if the absorbency of the trap tends to one, the index of global intransitiveness of the system tends to $-\frac{1}{2}$.

Based on the previous arguments, if we have a distribution of probability equally concentrated on two nodes of the cycle at first, we always obtain $\sigma = 0$. Indeed, placing the traps in those nodes, the casual agents who move on the network are equally shared between the two traps.

We can also study the competition on the cycle introducing two directions of travel for the walkers, with a symmetry parameter $p \in [0,1]$. For this purpose, we modify the stochastic matrix $\mathbf{M}$ of the system in the following way,
\begin{equation}
@{M}_{p} = (1-p)@{M} +p@{M}^T.    
\end{equation}

When $p = 0$ (or $p =1$) we still have a directed cycle only viable in one direction. If $p = \frac{1}{2}$, no direction of travel prevails in the system and $\mathbf{M}_{p}$ is symmetric. While values of $p$ between zero and one half introduce a drift in the system, i.e. the casual agents statistically move clockwise (or anticlockwise).

Let us study the competition on the cycle varying the symmetry parameter $p$, assuming fully absorbing traps and an initial uniform distribution of probability of the walkers on the network (Fig.~\ref{fig:distr}-a).
We observe that, independently on its size, the system is globally intransitive $\forall \, p$.

In addition, we note that, when $p = \frac{1}{2}$, $\sigma = 0$: this result was expected, because a symmetric $\mathbf{M}$ involves that all the possible positions of the traps on the cycle are equivalent. Then, the increase of the size of the system induces in turn an increase of intransitiveness degree: if the network allows only one direction of travel to the walkers, $\sigma \rightarrow \frac{1}{2}$, coherently with the behaviour of the index of global intransitiveness shown in Fig.~\ref{fig:distr}-a.

Assuming different initial distribution of probability for the walkers, the variation of $p$ allows to show non-trivial intransitiveness zones. This is the case for instance of an initial step-like distribution of probability (Fig.~\ref{fig:distr}-d), in which the casual walkers are initially focused in a fixed number of adjacent nodes. Such as in the case of the variation of the absorbency $\eps$ of the traps, if the step extends over one or two nodes, respectively $\sigma = -\frac{1}{2}$ or $\sigma = 0$ $\forall \, \eta$. 

The situation becomes more interesting if the step extends at least over 3 nodes (Fig.~\ref{fig:distr}-d). We observe a zone of transitivity around $p = \frac{1}{2}$, that becomes smaller to the increase of the step's width, until it disappears completely when the step-like distribution becomes an uniform distribution of probability. In these cases, there is a dominant position where the trap can be placed. 

In particular, we note that the the system is transitive when $p = \frac{1}{2}$ and so $\mathbf{M}$ is symmetric: this fact suggests that the initial distribution of probability of the casual walkers on the network generally plays a key role for the intransitiveness of competition between absorbing traps.

\subsection{Competition on the World Wide Web}

A meaningful and more complicated applied example where it is important to devise optimal navigation strategies is represented by the World Wide Web. A web surfer crawling the World Wide Web (WWW) network going from one site to another one by following the directed links on the different pages is a perfect Example of random walker on a complex network. Specific goals make us devise optimized searching schemes for target nodes.

Empirical observations of the Internet network have given us the knowledge of the specific topological rules that this graph undergoes. It is in particular clear that it is a directed network and that its degree distribution is scale-free~\cite{scalefree, www}. Moreover, in real WWW networks, there is a large number of "leaves", pages with links but which are not reachable since no other page contains a link to them. 

Thus, in order to reproduce the web surfing process, we  analysed a network of 100 nodes which has been generated by using the Barabasi-Albert algorithm~\cite{BarabasiAlbert}, studying  its intransitiveness at varying the absorbency index $\eps$. As shown in Fig.~\ref{fig:net}-a, in this case we have an intransitive region for intermediate values of the absorbency index $\eps$. By visually inspecting the scale-free network, we have found that this behaviour is given by structures like that reported in Fig.~\ref{fig:net}-b, where the large number of leaves  may make it intransitive for a partial absorbency of traps. This is indeed the case, as reported in Fig.~\ref{fig:net}-c.

\begin{figure}[t]
\centering
\begin{tabular}{cc}
  \begin{minipage}[b]{0.45\columnwidth}
    \scriptsize
    \setcounter{MaxMatrixCols}{12}
    $I = \begin{pmatrix}
      1&1&2&2&2&2&3&3&3&3&3&3\\
      1&1&2&2&2&2&3&3&3&3&3&3\\
      2&2&1&1&2&2&3&3&3&3&3&3\\
      2&2&1&1&2&2&3&3&3&3&3&3\\
      2&2&2&2&1&1&3&3&3&3&3&3\\
      2&2&2&2&1&1&3&3&3&3&3&3\\
      3&3&3&3&3&3&1&1&2&2&2&2\\
      3&3&3&3&3&3&1&1&2&2&2&2\\
      3&3&3&3&3&3&2&2&1&1&2&2\\
      3&3&3&3&3&3&2&2&1&1&2&2\\
      3&3&3&3&3&3&2&2&2&2&1&1\\
      3&3&3&3&3&3&2&2&2&2&1&1
\end{pmatrix}$\vspace{1cm}\\
  \end{minipage} & 
\begin{minipage}[t]{0.45\columnwidth}
  \includegraphics[width=\linewidth]{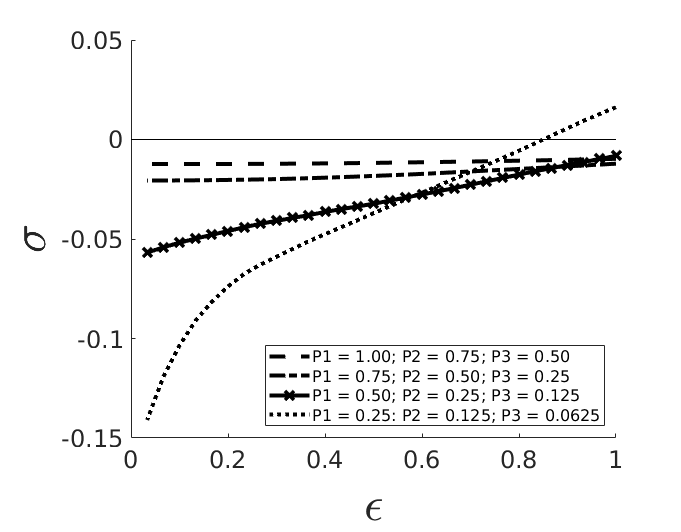}\end{minipage}
\end{tabular}
\caption{
(a) Example index matrix for hierarchical matrices (communities $l_1=2, l_2=6, l_3=2$): for an entry $k=I_{ij}$, the corresponding adjacency matrix has ones with probability $p_k$. (b) Average intransitiveness index $\sigma$ vs absorbency $\epsilon$ for hierarchical communities ($l_1=l_2=l_3=3$, average over 1000 simulations).}
\label{fig:hierarchical}
\end{figure}

\subsection{Hierarchical matrices}
We have also analysed a hierarchical matrix, which can well represent a model for communities. The network is defined by blocks, as indicated in Fig.~\ref{fig:hierarchical}-a~\cite{girvan2002community}. In each element of a block $k$, of size $l_k$, there is a one (connection) with probability $p_k$. As shown in Fig.~\ref{fig:hierarchical}-b, if communities are well isolated ($p_3\simeq 0$) and for enough absorbing traps, there is a region of intransitiveness. 

\begin{figure}
  \centering
  \begin{tabular}{cc}
  \includegraphics[width=0.3\linewidth]{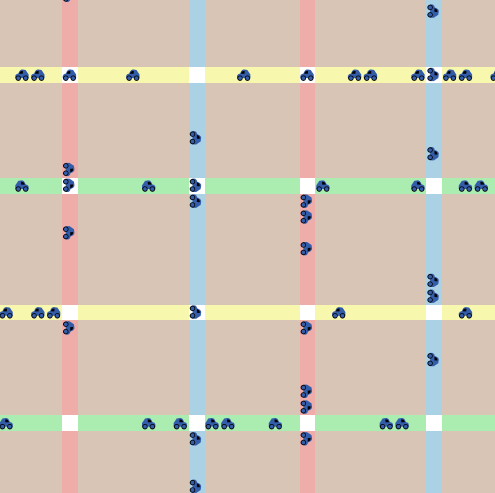} & 
  \includegraphics[width=0.45\linewidth]{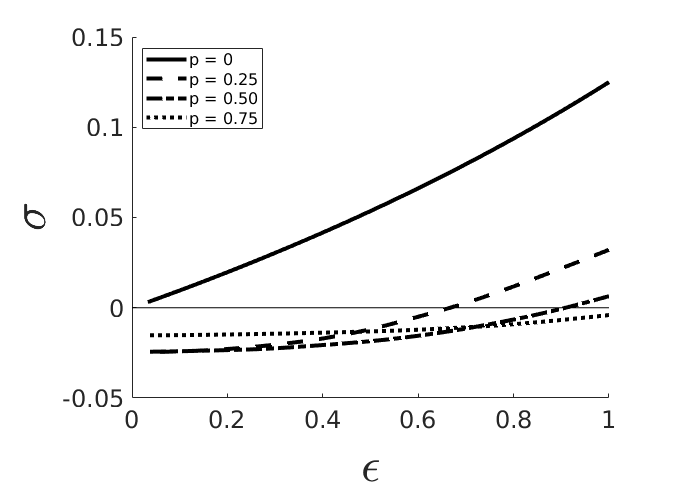}\\
  (a) & (b) \\
  \end{tabular}
  \caption{
    (a) A ``square'' city, where streets are alternating one-way ($p=0$). (b) Average intransitiveness index $\sigma$ vs absorbency $\epsilon$ for a ``square'' city, here $p$ denotes the probability of having two-way streets instead of alternating one-way (city with $4\times4$ streets, average over 1000 simulations.}
  \label{fig:squarecity}
\end{figure}

\subsection{Squarecity}
Finally, we examined the case of a ``square'' city, Fig.~\ref{fig:squarecity}-a, with streets that have a probability $p$ of being two-ways, and $1-p$ of being one-way in alternation. For small enough $p$, and large enough absorbency we have intransitiveness, Fig.~\ref{fig:squarecity}-b, again stating the importance of cycles.

\section{Conclusions}

We  introduced the concept of intransitiveness for games, which is the condition for which there is no first-player winning strategy, and the second player can statistically win. Since a game can be converted into a random walk on a graph, we can extend the intransitiveness concept to Markov processes, but this property depends also on the initial distribution of probability. 

In the case of random walks and Markov processes, the condition corresponding to the end of a game is that of walkers falling on a trap. Therefore, we have studied the competition among traps on random walks. 

We analyse in details this problem for the Penney game (an extension of the heads ot tails game which is intransitive for sequences longer than three), for walks on a circle and for a scale-free network, reminiscent of the structure of the world wide web. We also introduce several variations: traps can be partially absorbing, the walk can be biased and the initial distribution can be arbitrary. 

We found that the intransitiveness concept can be quite useful for characterizing the properties of a graph and of a given dynamics, like that of pages in the Word Wide Web, hierarchical communities and one-way streets in cities.


\section*{acknowledgments} The present version is an extension of the article of Ref.~\cite{baldi2019}, with the consensus of the other authors of the previous paper. In this version we added the part about traps with different absorbency in the Penney game, the dynamics in communities and in the square city.


\bibliographystyle{unsrtnat}
\bibliography{references}
\end{document}